\documentclass[vecphys]{svmult}

\usepackage{makeidx}         
\usepackage{graphicx}        
\usepackage{multicol}        
\usepackage[bottom]{footmisc}

\makeindex             

\begin{document}

\title*{Short-Period Variables in the Local Group Dwarf Galaxies Tucana and LGS3}

\titlerunning{Short-Period Variables in Tucana and LGS3}


\author{E. J. Bernard\inst{1}, M. Monelli\inst{1}, C. Gallart\inst{1}
\and the LCID Team: A. Aparicio\inst{1}, G. Bertelli\inst{2},
S. Cassisi\inst{3}, A. A. Cole\inst{4}, P. Demarque\inst{5},
A. E. Dolphin\inst{6}, I. Drozdovsky\inst{1}, H. C. Ferguson\inst{7},
S. Hidalgo\inst{4}, M. Mateo\inst{8}, L. Mayer\inst{9,10},
J. Navarro\inst{11}, F. Pont\inst{12}, E. D. Skillman\inst{4},
P. B. Stetson\inst{13}, E. Tolstoy\inst{14}}

\institute{Instituto de Astrof\'isica de Canarias, E-38205 La Laguna,
 Tenerife, Spain
\texttt{ebernard@iac.es}
\and INAF, Osservatorio Astronomico di Padova, Italy
\and INAF, Osservatorio Astronomico di Teramo, Italy
\and University of Minnesota, USA
\and Yale University, USA
\and University of Arizona, USA
\and STScI, USA
\and University of Michigan, USA
\and Institut f\"ur Theoretische Physik, University of Z\"urich, Switzerland
\and Institut f\"ur Astronomie, Physics Department, ETH Zurich, Switzerland
\and University of Victoria, Canada
\and Observatoire de Gen\`eve, Switzerland
\and Dominion Astrophysical Observatory, Canada
\and Kapteyn Astronomical Institute, Netherlands}

\authorrunning{Edouard J. Bernard, Matteo Monelli, Carme Gallart and the LCID Team}

%
%
\maketitle
\index{Edouard J. Bernard}

\begin{abstract}
 We present preliminary results concerning the search for short-period
 variable stars in Tucana and LGS3 based on very deep HST/ACS imaging. 
 In the fraction of the observed field we studied in each galaxy, a total of
 133 and 30 variables were found, respectively. 
 For Tucana, we identified 76 of them as RR Lyrae (RRL) stars pulsating in the
 fundamental mode (RR$ab$) and 32 in the first-overtone mode (RR$c$), as well
 as 2 anomalous Cepheids (AC).
 The mean period of the RR$ab$ and RR$c$ is 0.59 and 0.35 days, respectively.
 In the case of LGS3, we found 24 RR$ab$ and 4 RR$c$, with mean periods 0.61
 and 0.39 days, respectively, plus two candidate ACs.
 These values place both galaxies in the Oosterhoff gap.
\end{abstract}

\section{Introduction} \label{sec:1}

 Pulsating variable stars play a fundamental role in the study of stellar
 populations and in cosmology. In particular, classical Cepheids (CC) and
 RRL stars are primary distance indicators and have a fundamental importance
 in the calibration of the extragalactic distance scale through their
 period-luminosity relation.

\begin{figure}
\centering
\includegraphics[height=6.4cm]{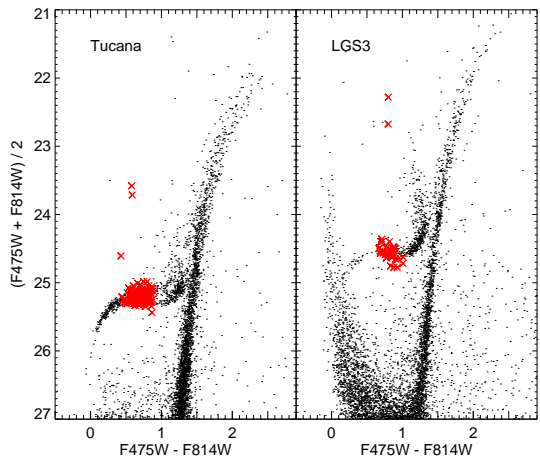}
\caption{Color-magnitude diagrams of the dSph Tucana ({\it left}) and the
 transition dSph/dIrr LGS3 ({\it right}). Small dots represent the stars
 from the whole observed field, while the studied candidate variables are
 shown as red crosses.}
\label{fig:1}
\end{figure}
 
 The pulsational properties of RRL stars are also used to estimate the
 local reddening (\cite{cle04}), and the metallicity retrieved
 from the empirical relation between period and amplitude (\cite{san04})
 was found to be in good agreement with the mean
 abundance from the RGB (\cite{pri05}). Moreover, theoretical
 investigations indicate that the pulsation characteristics of RRL
 stars -- topology of the instability strip, period and amplitude
 distribution -- can be safely adopted to constrain the properties of the
 parent population (age, metallicity, horizontal branch (HB) morphology) as
 well as to put tight constraints on the mass, effective temperature, and
 distance modulus of the individual stars (\cite{bon97}, \cite{bon00}).

 RRL stars, ACs and CC allow us tracing of the old (t$>$10\ Gyr),
 intermediate age (a few Gyr)
 and young ($<$100\ Myr) stellar populations, respectively, therefore
 highlighting the eventual radial trends across the studied galaxy
 (eg. Phoenix: \cite{gal04}, Leo I: \cite{bal04}). This, in turn, hints on
 the star formation history and the formation mechanisms of the host galaxy.
 
\begin{figure}
\centering
\includegraphics[height=6.6cm]{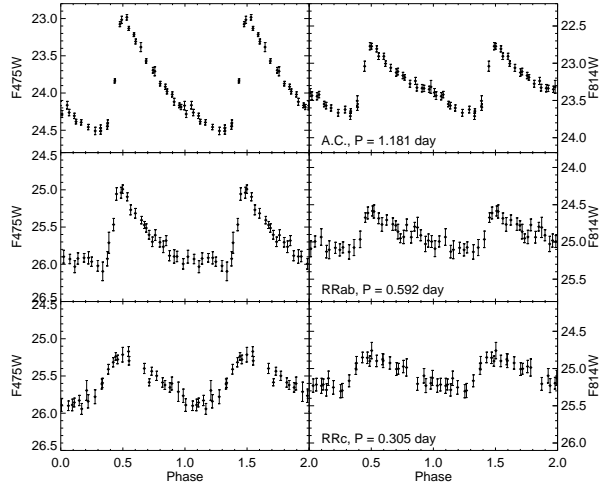}
\caption{From top to bottom, the light-curves of typical anomalous Cepheid, 
 RR$ab$ and RR$c$ in the F475W ({\it left}) and F814W ({\it right}) bands
 for Tucana.}
\label{fig:2}
\end{figure}

\section{The Data}
 
 Until recently, the study of faint variables like RRL stars was limited to
 the stellar systems associated to the Milky Way: its halo, globular clusters
 (GCs) and satellite dwarf spheroidal galaxies (dSph). The launch of the
 HST opened the possibility to obtain resolved-star photometry reaching
 the HB of crowded field to greater distances. The RRL stars of the Andromeda
 galaxy and its vicinity have now been extensively surveyed (eg. \cite{bro04}, 
 \cite{ric05}). For the first time, as part of the cycle 14 HST/ACS program
 `The Onset of Star Formation in the Universe: Constraints from Nearby Isolated
 Dwarf Galaxies', we obtained very deep ($V\sim29$) multi-epoch images of four
 {\it isolated} dwarf galaxies of the Local Group: the dwarf spheroidals Tucana
 and Cetus, the dwarf irregular (dIrr) IC1613 and the so-called transition
 dSph/dIrr LGS3.

 The DAOPHOT-II/ALLSTAR and ALLFRAME packages (\cite{ste94})
 were used to obtain the instrumental photometry of the stars on the individual
 images. Figure~\ref{fig:1} presents the resulting color-magnitude diagrams for
 Tucana and LGS3. The candidate variables were extracted from the star list
 using the variability index given by DAOMASTER.
 Period search for the candidates was done using an implementation of the phase
 dispersion minimization method (\cite{ste78}) taking into account the
 information from both bands simultaneously. A sample of the light-curves for
 Tucana variables is presented Fig.~\ref{fig:2}.

\section{Results}

 As a preliminary work, we focused on the candidate variables found on one half
 of the ACS chip 1. This strip of $3' \times 0.75'$, closest to chip 2, samples
 a somewhat radial portion of the galaxies. It prevents the bias that a
 gradient in the properties of the variables could have on the average values.
  
\begin{figure}
\centering
\includegraphics[width=9.61cm]{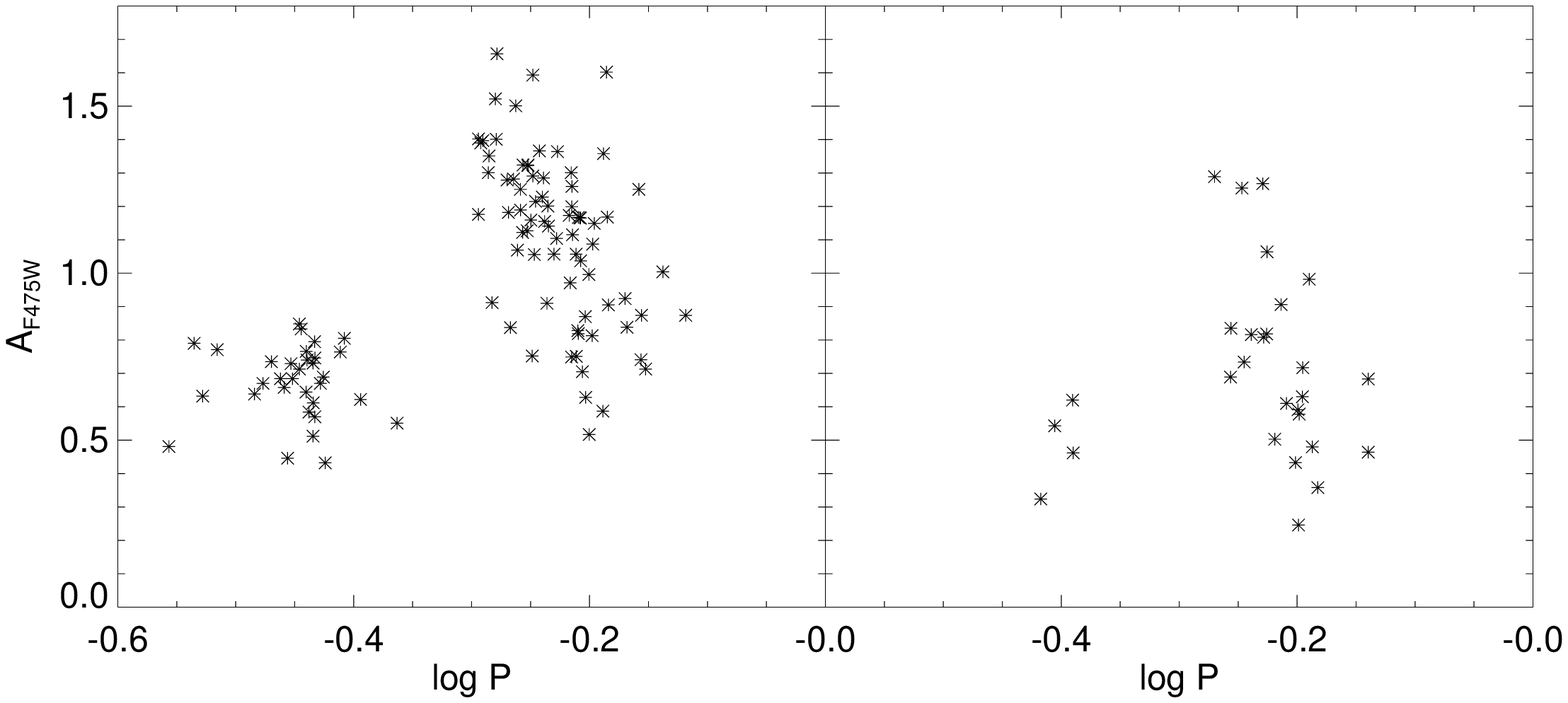}
\caption{Period-amplitude diagrams for the RRLs in Tucana ({\it left}) and
 LGS3 ({\it right}).}
\label{fig:3}
\end{figure}
 
 In the studied field of Tucana, we found 76 RR$ab$ and 32 RR$c$ with mean
 periods of 0.59 and 0.35 days, respectively. The period is ill-defined for
 another 23 HB candidates. Some of these candidates could be double-mode RRL
 (RR$d$). Using only the RRL stars with well-defined periods, we find the ratio
 of RR$c$ to the total number of RRL stars to be 0.30. 
 The two variables located $\sim$1.5 magnitude above the HB are most likely
 ACs: they have well defined light-curves with large amplitude (A$_{F475W}
 \sim 1.5$) and $p\sim1.14$ days.

 For LGS3, the small number of datapoints (12 in each band, vs. 32 for Tucana)
 made uncertain the period estimates,
 and the particular temporal sampling created strong aliasing. However, the
 period-amplitude diagram presented in Fig.~\ref{fig:3} seems to support our
 choice of the period. It also shows that the amplitude of the RRL stars is
 systematically smaller in LGS3 than in Tucana. Although some of the lowest
 amplitudes are due to the lack of observations at maximum light, this trend
 has been noted for dSph and GC having a very red HB (see \cite{pri05} and
 references therein). Indeed, the HB of LGS3 is mainly red, with RRL stars 
 preferentially located near the red edge of the instability strip.

 With the metallicities [Fe/H]=-1.5 and [Fe/H]=-1.8 obtained by fitting
 isochrones on the CMDs of Tucana and LGS3, respectively, we found their mean
 RR$ab$ period to be in excellent agreement with the mean period-metallicity
 relation of the GCs and dSph of the Milky Way. These values place both
 galaxies in the Oosterhoff-Intermediate group, as almost all the Milky Way
 dSph companions (\cite{cat04}).

%



\printindex
\end{document}